\documentstyle[12pt]{article}

\begin{document}
\pagestyle{plain}

\newcommand{\vp}{\varphi}
\newcommand{\pr}{\prime}
\newcommand{\be}{\begin{equation}}
\newcommand{\ee}{\end{equation}}
\newcommand{\bea}{\begin{eqnarray}}
\newcommand{\eea}{\end{eqnarray}}

%
\title{\bf SUSY-BASED VARIATIONAL METHOD FOR THE ANHARMONIC OSCILLATOR}
\author{
Fred Cooper \\
{\small \sl Theoretical Division, Los Alamos National Laboratory,}\\
{\small \sl Los Alamos, NM 87545}\\
{\small \sl and   }\\
{\small \sl Physics Department, University of New Hampshire,} \\
{\small \sl Durham, NH 03824}\\
John Dawson and Harvey Shepard \\
{\small \sl Physics Department, University of New Hampshire,} \\
{\small \sl Durham, NH 03824}\\ \\
       }
\maketitle

\begin{abstract}
Using a newly suggested algorithm of Gozzi, Reuter,
and Thacker for calculating the excited states of one
dimensional systems, we determine approximately the eigenvalues
 and eigenfunctions of the
anharmonic oscillator, described by the Hamiltonian $H= p^2/2 + g x^4$.
We use ground state post-gaussian trial wave functions of the form
$\Psi(x) = N{\rm{exp}}[-b |x|^{2n}]$, where $n$ and $b$ are continuous
variational parameters.  This algorithm is based on the hierarchy of
 Hamiltonians related
by supersymmetry (SUSY) and the factorization method.  We find that our two
parameter family of trial wave functions yields excellent energy eigenvalues
and wave functions for the first few levels of the anharmonic oscillator.
\end{abstract}

The central idea of the factorization method for ordinary differential
equations, introduced by Schr\"{o}dinger \cite{Schrodinger} and by Infeld and
Hull \cite{Infeld} to solve problems in quantum mechanics, is the recognition
that once the ground state energy and wave function of a one-dimensional
potential problem are known, then the potential is determined,
as well as the factors of the Hamiltonian.
Recently, Gozzi, Reuter, and Thacker \cite{Gozzi} proposed a simple strategy
for solving such one-dimensional potential problems approximately in a
variational scheme, based on utilizing the hierarchy of Hamiltonians that are
related by supersymmetry (SUSY) and factorization
\cite{Sukumar,Adrianov,others}.
We apply their method here to the case of a one-dimensional anharmonic
oscillator, using a simple variational wave function, and examine the
results.

First let us review briefly the formalism for describing the hierarchy of
isospectral Hamiltonians. Consider the Hamiltonian $H_1$
having $k$ bound states whose discrete eigenvalues and eigenvectors are denoted
by $E_n^{(1)}$ and $\Psi_n^{(1)}$. Once we subtract the ground state
energy, we can factorize the Hamiltonian as follows:

\begin{eqnarray}
   H_1 -E_0^{(1)}  =  {1 \over 2} A_1^{\dag} A_1 & \equiv &
 {1 \over 2} [-{d \over
dx}+ W_1(x)] [{d \over dx} + W_1(x)] \nonumber \\
   & = & -{1 \over 2}{d^2 \over dx^2} + V_1(x) \, ,
\end{eqnarray}
where the superpotential,
\be
   W_1(x) = - {d \over dx} [\ln \Psi_0^{(1)}(x) ],\nonumber
\ee
and
\be
V_1 \equiv (1/2) (W_1^2 - W_1^{'}) \>. \nonumber
\ee
The second Hamiltonian $H_2$ in the hierarchy, the SUSY partner of
$H_1$,
\be
  H_{2} = {1 \over 2} A_1 A_1^{\dag}+E_0^{(1)}
\ee
has the same spectrum as $H_1$, except that there is no state in
 $H_2$ corresponding to
the ground state of $H_1$.
 Introduce the notation that in $E_n^{(m)}$, $n$ labels the energy
level and $(m)$ refers to the m-th Hamiltonian $H_m$.

One can show \cite{Witten} \cite{coop83} that the energy eigenvalues and
eigenfunctions of the two  Hamiltonians $H_1$ and $H_2$ are related by
\[
E_{n+1} ^{(1)} = E_n^{(2)}, \hspace{.3in} \Psi_n^{(2)} =
2^{-1/2}(E_{n+1} ^{(1)} - E_0^{(1)})^{-1/2} A_1 \Psi_{n+1}^{(1)}
\]
\be
 \Psi_{n+1}^{(1)} =
2^{-1/2} (E_{n+1} ^{(1)} - E_0^{(1)})^{-1/2} A_1^{\dag} \Psi_{n}^{(2)}
\ee
and in particular:
 \be
     (-{d \over dx} + W_1 )  \Psi_0^{(2)} = 2^{1/2}(E_{1} ^{(1)} -
E_0^{(1)})^{1/2}
\Psi_{1}^{(1)}
 \ee
so that the first excited state of the original Hamiltonian $H_1$
 can be obtained from
the ground state of the supersymmetric partner potential.
If we determine the
ground state wave function $\Psi_0^{(2)}$ and ground state energy
(=$E_1^{(1)}$)  for $H_{2}$,
then we can repeat the above argument and determine a third Hamiltonian $H_3$
as the SUSY
partner of $H_2$. Namely  we can write
\be
  H_{2}=  {1 \over 2} A_1 A_1^{\dag}+E_0^{(1)}=
  {1 \over 2} A_2^{\dag} A_2+E_1^{(1)} \nonumber
 \ee
where
\be
A_2 = {d \over dx} + W_2(x).
\ee
Then
\begin{equation}
  H_{3}=  {1 \over 2} A_2 A_2^{\dag} + E_1^{(1)}  \>.
\end{equation}
We also notice that the Hamiltonian given by $\bar{H}_2= {1 \over 2} A_1
A_1^{\dag}$ has a ground state energy which is the energy difference
$E_1^{(1)}-E_0^{(1)}$. Furthermore we have the relationships:
\be
E_{n+2} ^{(1)} = E_{n+1} ^{(2)}=E_n^{(3)},\nonumber
\ee
\begin{eqnarray}
  \Psi_{n+1}^{(1)} & = & 2^{-1/2} (E_{n+1} ^{(1)} - E_0^{(1)})^{-1/2}
 A_1^{\dag} \Psi_{n}^{(2)}\nonumber \\
 & = & (1/2) (E_{n+1} ^{(1)} - E_0^{(1)})^{-1/2}
 (E_{n+1} ^{(1)} - E_1^{(1)})^{-1/2}  A_1^{\dag}  A_2^{\dag}\Psi_{n-1}^{(3)}.
\end{eqnarray}
In particular, for $n=1$ we determine the second excited state
 of $H_1$ from the
ground state of $H_3$. Clearly this method can be generalized
 to obtain $k$ Hamiltonians
of the hierarchy, where k is the total number of bound states of $H_1$.

The scheme of Gozzi {\em et.~al.} \cite{Gozzi} is to determine the ground state
wave function
$\Psi_{0}^{(1)}$ and
energy $E_0^{(1)}$ of the original Hamiltonian $H_1$ using a variational
method, then to approximately find the  superpotential
\begin{equation}
   W_{1\,v}(x) = - \frac{d}{dx} \Bigl\{ \ln \Psi_{0\,v}^{(1)}(x) \Bigr\}  \>.
\end{equation}
This allows one to approximately determine the first SUSY
 partner potential
corresponding to $\bar{H}_2$
\begin{equation}
   V_{2\,v}(x) = \frac{1}{2} \Bigl[ W^2_{1\,v}(x) + W'_{1\,v}(x) \Bigr]
\end{equation}
obtained from the approximate partner Hamiltonian $\bar{H}_{2\,v}$
\begin{equation}
   \bar{H}_{2\,v} = \frac{1}{2} A_{1\,v}^{\vphantom{\dagger}}
A_{1\,v}^{\dagger},\hspace{.2in} A_{1\,v} ={d \over dx} + W_{1\,v} \>.
\end{equation}
whose ground state energy determines approximately the energy difference
$E_1^{(1)} -E_0^{(1)}$ as discuused earlier.
One  next finds the variational ground state wave function and energy of the
approximate SUSY partner Hamiltonian $H_{2\,v}$ from which one can also
calculate
the approximate first  excited state of the original Hamiltonian, and we can
proceed as described above, using a new variational approximation for each
ground state, until all the wave  functions and energies of this approximate
hierarchy of Hamiltonians are known.

The interesting question is how accurate is such a scheme?  All the
Hamiltonians of
the variational hierarchy are not the real Hamiltonians of the
hierarchy but approximate
ones related to the variational approximation of the ground state wave
function of the
original problem.  Thus we do not {\em a priori} know that we are calculating
accurately the energy eigenvalues of the original Hamiltonian.
 We only know that we are accurately determining
the energies of the   approximate  Hamiltonians of the hierarchy.
Thus, for this scheme to work, one needs an extremely
 accurate method of finding
the ground state wave function and energy of $H_1$. Also in this scheme the
higher wave functions of $H_1$
are obtained by repeated derivatives of the ground state wave functions of the
approximate hamiltonians. Thus we expect a loss in accuracy as we go to the
higher wave functions.

In some recent work on soliton dynamics in various nonlinear systems including
the
nonlinear Schr\"{o}dinger equation \cite{CSLS}, we have demonstrated the
efficacy of
post-gaussian time dependent variational wave functions of the form
\begin{equation}
u(x,t) = N(t) \, \exp [-b(t) |x-q(t)|^{2n}] \>.
\end{equation}
These wave functions, when used to minimize the action, gave excellent
global fits to
various solitons as well as
 very accurate determinations of the soliton energies.
We found that these wave functions are
globally accurate to a
few percent or so, and give energy eigenvalues accurate to about $0.1$\%.

Thus, we propose to use as our trial wave functions for the ground
state wave functions of the hierarchy of Hamiltonians,
\begin{equation}
     \Psi_{0}^{(k)} = N_k \, \exp \Bigl[
        - \frac{1}{2} \Bigl| \frac{x^2}{\rho_k} \Bigr|^{n_k} \Bigr]
     \>, \hspace{.5in}
     N_k = \Bigl[ 2 \sqrt{\rho_k} \;\Gamma( 1 + \frac{1}{2 n_k} ) \Bigr]^{-1/2}
     \>.
\label{eq:trialwf}
\end{equation}

We first scale the Hamiltonian for the anharmonic oscillator,
\begin{equation}
  H = - \frac{1}{2} \frac{d^2}{dx^2} + g x^4  \>,
\end{equation}
by letting $x \rightarrow x/g^{1/6}$ and $H \rightarrow g^{1/3} H$.
Then we find the ground state energy of the anharmonic oscillator and the
variational parameters $\rho_1$ and $n_1$ by forming the functional
\begin{equation}
     E_0(\rho_1,n_1) =
        \langle \, \Psi_{0} | - \frac{1}{2} \frac{d^2}{dx^2} +   x^4 |
        \Psi_{0} \, \rangle \, .
\end{equation}
We determine $\rho_1$ and $n_1$ by requiring
\begin{equation}
     {\partial E_0 \over \partial \rho_1} = 0  \>, \hspace{.5in}
   {\partial E_0 \over \partial n_1} = 0  \>.
\end{equation}
The equation for the energy functional for the anharmonic oscillator is:
\begin{equation}
   E_0(\rho_1,n_1) =
     { n_1^2 \over \ 2 \rho_1 }
     { \Gamma( 2-{1 \over 2n_1}  ) \over \Gamma( { 1 \over 2n_1 } ) } +
     \rho_1^2
     { \Gamma( { 5 \over 2n_1 } ) \over \Gamma( { 1 \over 2n_1 } ) }\, .
\end{equation}
Minimizing this expression, we obtain the variational result
\begin{equation}
  E_0 = 0.66933, \hspace{.2in} n_1=1.18346, \hspace{.2in} \rho_1 =
  0.666721 \, .
\end{equation}
This ground state energy is to be compared with a numerical evaluation which
yields
$0.667986$, which is of the accuracy obtained in our soliton calculations.
Since the trial wave function for all ground states is given by
(\ref{eq:trialwf}), the variational superpotential for all $k$ is
\begin{equation}
   W_{k\,v} = n_k |x|^{2n_k -1} (\rho_k)^{-n_k}  \>.
\end{equation}
Since we are interested in the energy differences $ E_k - E_{k-1}$  of the
anharmonic oscillator we will concentrate on the variational Hamiltonian
\be
   \bar{H}_{k+1\,v} =  \frac{1}{2} A_{k\,v} A_{k\,v}^{\dagger}
\ee
which approximately determines these energy differences as discussed above.

First, let us  compare the exact potential $ x^4- E_0$ with the
variational one
obtained using the ground state variational wave function:
\begin{equation}
V_{1\,v}(x) = \frac{1}{2} \Bigl[ W^2_{1\,v}(x) - W'_{1\,v}(x) \Bigr]  \>.
\end{equation}
This is shown in the lower two curves of Fig.~1 for the most relevant
 range  $0 \leq x \leq 1.5$ for
determining the ground state wave function.  We see that the variational
potential
is quite different from the exact one, although it has almost the same ground
state energy.  It is also interesting to  compare our variational
``partner potential,''
\begin{equation}
V_{2\,v}(x) = \frac{1}{2} \Bigl[ W^2_{1\,v}(x) + W'_{1\,v}(x) \Bigr]  \>,
\end{equation}
with the exact one, obtained by first solving the Riccati equation for $W_1$,
\begin{equation}
   \frac{1}{2} \Bigl[ W^2_1(x) - W'_1(x) \Bigr] = x^4 - E_0 \>,
\end{equation}
or equivalently solving the Schr\"{o}dinger equation for
\begin{equation}
   \Bigl[ - \frac{1}{2} \frac{d^2}{dx^2} + x^4 \Bigr] \, \Psi_0 = E_0 \Psi_0
\end{equation}
for $\Psi_0$, and determing $W_1$ from the logarithmic derivative of $\Psi_0$.
The result of this calculation of the exact partner potential
 and its variational
approximation are shown in the upper two curves of Fig.~1.

We obtain the approximate energy splittings by minimizing the energy functional
\begin{equation}
  \Delta E_k (\rho_k,n_k) = \frac{1}{2}
     \langle \, \Psi_{0\,v}^{(k+1)} | \frac{d^2}{dx^2} +
            W^2_{k\,v} + W'_{k\,v} | \Psi_{0\,v}^{(k+1)}
\, \rangle =E^{(1)}_k - E^{(1)}_{k-1}  \>.
\end{equation}
These integrals are trivial to carry out, and one obtains the simple recursion
relation
\[
 \Delta E_k (\rho_k,n_k) =
   \frac{ n_k^2 }{ 2 \rho_k }
   \frac{ \Gamma\left( 2-{ 1 \over 2n_k } \right) }
        { \Gamma\left( { 1 \over 2n_k } \right) } +
   \frac{ n_{k-1}^2 }{ 2 \rho_k }
   \left( \frac{ \rho_k }{ \rho_{k-1} } \right)^{ 2n_{k-1} }
   \frac{ \Gamma\left( { 4n_{k-1}-1 \over 2n_k } \right) }
        { \Gamma\left( { 1 \over 2n_k } \right) }
\]
\begin{equation}
   + \frac{ n_{k-1} }{ 2 \rho_{k} }  ( 2n_{k-1} - 1 )
     \left( \frac{ \rho_k }{ \rho_{k-1} } \right)^{ n_{k-1} }
     \frac{ \Gamma\left( \frac{ 2n_{k-1}-1 }{ 2n_k } \right) }
          { \Gamma\left( \frac{ 1 }{ 2n_k } \right) }\, .
\end{equation}
One can perform the minimization in $\rho$ analytically leaving one
minimization
to perform numerically.

The results for the variational parameters and for the energy
differences are presented
in Table 1 for the first five energy eigenvalues and compared with our
numerical
calculation, based on a shooting method, which essentially agrees with the
results of
Hioe and Montroll\cite{Hioe}, who made an exhaustive study of the
anharmonic oscillator.
For the ground state and first two excited states, we get excellent agreement
with the numerical results.
However, for the third excited state and higher, the variational approach
relaxes
to a Gaussian which
leads to equal spacing with a $\Delta E$  of around $2.3$.
Thus for the higher states our method fails to give the correct large $n$
behavior,
which from WKB results, is known to be $1.376 ( n + \frac{1}{2})^{4/3}$.
So this method works  where WKB fails and {\em vice-versa}.
Of course, our results for the higher energy levels would be improved by using
more
variational parameters, which however would spoil the simplicity of
the calculation
presented here.

Finally, we have calculated the approximate excited states by repeated use of
the
approximate node insertion operators.  The results for the first three
variational wave functions are
compared to the exact numerical results in Fig 2.
We see that we are doing an excellent job in the regime $0 \leq x \leq
1.5$, which is
the most important region for determining the energy eigenvalues.
The higher eigenvalues require more and more derivatives, and thus the
accuracy of the
results diminish for large $x$ where the variational potential is not so
accurate.

%
%
%
\begin{center}
\begin{tabular}{| c | r@{.}l | r@{.}l | r@{.}l | r@{.}l |}
\multicolumn{9}{c}{Table 1: Variational Parameters and Energies} \\
\hline \hline
&
\multicolumn{2}{|c|}{} &
\multicolumn{2}{|c|}{} &
\multicolumn{4}{|c|}{$E_{n} - E_{n-1}$} \\
\cline{6-9}
\multicolumn{1}{|c|}{level} &
\multicolumn{2}{c|}{n} &
\multicolumn{2}{c|}{$\rho$} &
\multicolumn{2}{c|}{variation} &
\multicolumn{2}{c|}{exact}  \\
\hline
0 & 1&183458 & 0&666721 & 0&669330 & 0&667986 \\
1 & 0&995834 & 0&429829 & 1&727582 & 1&725658 \\
2 & 1&000596 & 0&435604 & 2&316410 & 2&303151 \\
3 & 0&999917 & 0&434779 & 2&297082 & 2&638935 \\
4 & 1&000012 & 0&434894 & 2&299820 & 2&908578 \\
\hline \hline
\end{tabular}
\end{center}
\bigskip

\vspace{1.0in}
\noindent{\bf FIGURE CAPTIONS}\\

\vspace{0.2in}

\noindent Fig. 1: Exact (solid line) and variational (dashed line)
potentials for the anharmonic oscillator (lower curves) and its SUSY
partner (upper curves).

\vspace{0.1in}

\noindent Fig. 2: Exact (solid line) and variational (dashed line)
wave functions for the lowest three levels of the anharmonic
oscillator.

\end{document}